\documentclass[12pt]{article}
\usepackage{graphicx}
\title{Spectrum of $^{10}$Be in the basis of the leading representation of the SU(3) group}
\author{G.~F.~Filippov$^{1)}$, K.~Kat\=o$^{2)}$, S.~V.~Korennov$^{1,2)}$ and
Yu.~A.~Lashko$^{1)}$\\
\normalsize $^1$ Bogolyubov Institute for Theoretical Physics, Kiev, Ukraine\\
\normalsize $^2$ Graduate School of Science, Hokkaido University,
Sapporo, Japan}
\date{}
\begin{document}

\maketitle

\begin{abstract}

A realization of the approximation of the SU(3) leading
representation with the microscopic Hamiltonian and a
nucleon-nucleon interaction is presented in detail. An effective
Hamiltonian reproducing results of calculations with some known
potentials is constructed. It is shown that its structure is quite
similar to that of the triaxial rotator, and the wave functions in
the Elliott's scheme are linear combinations of Wigner's
$D$-functions although they should be properly normalized.

\end{abstract}

\section{Introduction}

For the theoretical studies of excitation spectra of valence
nucleons in light and medium nuclei, Elliott\cite{El} proposed the
basis states of irreducible representations $(\lambda,\mu)$ of the
SU(3) group satisfying the Pauli exclusion principle. A special
attention was paid to the most symmetric irreducible ("leading")
representations. In the states belonging to the leading
representations (LR) the number of even nucleonic pairs is the
largest, as they are characterized by an even orbital momentum of
the relative motion of the nucleons in the pairs, while the
interaction between these nucleons are reproduced by the
components $V_{2S+1=3,2T+1=1}$ and $V_{13}$ of a central
nucleon-nucleon (NN) potential. The number of odd nucleonic pairs
is, on the contrary, the smallest, the orbital momentum is odd,
and the acting components of a NN potential are $V_{33}$ and
$V_{11}$. While the even components are describing the strong
attraction between the nucleons, the odd ones are responsible for
the repulsion at short distances necessary to provide the
saturation of nucleon forces and the known dependence of the
volume of the nucleus on the number of nucleons. However, in the
Elliott's scheme, the main argument for the importance of the LR
was related not to the properties of the nuclear force, but rather
to the properties of the operator of quadrupole (QQ) interaction
$$-\eta{\bf QQ}=\hat{G}_2-{1\over2}{\bf L}^2,$$
which was chosen to describe the residual interaction in the
system. Here, $\eta$ is a positive parameter, $\hat{G}_2$ is the
second-order Casimir operator of the $su(3)$ algebra, and
$\hat{\bf L}^2$ is the squared orbital momentum. The eigenvalue
$g_2$ of the operator $\hat{G}_2$ is largest in the LR. As a
result, firstly, the minimal eigenvalue $-\eta{\bf QQ}$ is found
in the LR, and secondly, the model reproduces the ordering of
lowest levels of the principle rotational band.

The Elliott's scheme was further developed by the appearance of
the pseudo-SU(3) model\cite{HechtAr}, \cite{HechtAd},
\cite{HechtDr} and especially by the studies of the scissors
mode\cite{LoIud}. In these studies, it became necessary to improve
the phenomenological Hamiltonian and the wave functions of
even-even nuclei (the ground states and the final states of the
isovector M1 transitions)\cite{Dr}, \cite{Dr1}, \cite{Dr2}. One
more possible application of the SU(3) model is the neutron-rich
nuclei like $^{9,11}$Li and $^{10,11}$Be, where the calculations
may be made with some microscopic NN potential instead of just a
phenomenological Hamiltonian. One has to find such an effective
Hamiltonian expressed in terms of the SU(3) group generators,
which would generate the same spectrum as the microscopic
Hamiltonian. One has also to find those O(3)-scalar combinations
of the SU(3) generators which, in principle, can enter such a
Hamiltonian. This work deals with this kind of problems.

We shall see that the Elliott's scheme and the triaxial rotator
model share many common features. In particular, the basis states
in both models can be expressed in terms of the Wigner's
$D$-functions, but the normalization differs due to the fact that
the density matrix in the Elliott's scheme corresponds to a mixed
state, as opposed to a pure state in the case of the triaxial
rotator. This conclusion is supported by the structure of the
effective Hamiltonian in the Elliott's scheme. It has a form of a
linear combination of scalar expressions made of the SU(3) group
generators, and can be expressed through integer powers of the
Hamiltonian of the triaxial rotator. The larger the number of
quanta in the valence shell, the largest power of the rotator
Hamiltonian enters the Elliott's effective Hamiltonian.

We apply the LR approximation to the $^{10}$Be nucleus as an example.
The NN interaction is simulated by the known Volkov\cite{Vol} and
Minnesota\cite{Minn} potentials in order to test whether they can reproduce
the observed excitations of $^{10}$Be. Usually, the Elliott's scheme is used
with relatively simple phenomenological effective potentials. We therefore
show in detail the procedure in the case of a microscopic Hamiltonian
which, of course, is a generalization. We construct the scalar expressions
from the SU(3) group generators and show that their linear combination can
be reduced to the Hamiltonian of the triaxial rotator. Finally, we determine
the parameters of the phenomenological effective potential making it
equivalent to the microscopic one in the LR limit.

\section{Elliott's scheme and triaxial rotator}

Elliott's scheme (in its LR approximation) shares many common
features with the theory of the rigid triaxial rotator\cite{DF}.
It becomes clear in a space where the basis functions of both
models are represented by the spherical Wigner's functions. The
tool for the transition to such a space is the construction of the
function generating a complete LR basis in the form of a Slater
determinant. As a result, the matrix elements of various operators
are calculated easily. Operators involved are ${\bf LQL}$ and
${\bf LQQL}$ which have a simple algebraic interpretation\cite{BM}
and consisting of the SU(3) group generators, as well as
microscopic operators of central, spin-orbit and tensor NN
interactions.

Consider the case of the $^{10}$Be nucleus. The quantum numbers of basis
states are, in fact, known, and we just check that our approach generates
them correctly. At the same time, the explicit form of basis functions will
be established, along with the proper normalization, which is crucial in
microscopic calculations of spectra and transition probabilities.

We first define the orbitals, distinguishing between proton and
neutron configurations and restricting the basis with the minimal
allowed number of oscillator quanta. Below we omit the
spin-isospin quantum numbers for simplicity.

Two proton ($s-$ and $p-$) orbitals are
\begin{eqnarray}
\label{a1} \phi_{0\pi}({\bf
r})={1\over\pi^{3/4}}\exp\left(-{r^2\over2}\right)~\mbox{and}~
\phi_{1\pi}({\bf r})= {\sqrt{2}({\bf u}{\bf
r})\over\pi^{3/4}}\exp\left(-{r^2\over2}\right),
\end{eqnarray}
where the unit vector
${\bf u}$ is the first independent variable in the space where the Wigner's
functions will be defined later. There are two protons with different
spin projections in each of these two states.

The remaining six neutrons are allocated in pairs in three states,
\begin{eqnarray}
\label{a2} \phi_{0\nu}({\bf
r})={1\over\pi^{3/4}}\exp\left(-{r^2\over2}\right),~
\phi_{1\nu}({\bf r})= {\sqrt{2}({\bf u}{\bf
r})\over\pi^{3/4}}\exp\left(-{r^2\over2}\right),
\end{eqnarray}
\begin{eqnarray}
\label{a3} \phi_{2\nu}({\bf r})= {\sqrt{2}\over\pi^{3/4}}\cdot
{([{\bf w}\times\tilde{\bf u}]{\bf r})\over({\bf u}\tilde{\bf u})}
\exp\left(-{r^2\over2}\right).
\end{eqnarray}
There appeared another vector variable, the unit vector ${\bf w}$,
orthogonal to the vector ${\bf u}$. We also introduce the vectors
$\tilde{\bf u}$ and $\tilde{\bf w}$ for the conjugated orbitals.
These vectors are also mutually orthogonal, but, in general, have
a different orientation in space.

It is easy to see that
\begin{eqnarray*}
\int\phi^*_{0\pi}({\bf r})\phi_{0\pi}({\bf r})d{\bf r}=1,~
\int\phi^*_{1\pi}({\bf r})\phi_{1\pi}({\bf r})d{\bf r}=
({\bf u}\tilde{\bf u}),
\end{eqnarray*}
\begin{eqnarray*}
\int\phi^*_{0\nu}({\bf r})\phi_{0\nu}({\bf r})d{\bf r}=1,~
\int\phi^*_{1\nu}({\bf r})\phi_{1\nu}({\bf r})d{\bf r}=
({\bf u}\tilde{\bf u}),
\end{eqnarray*}
\begin{eqnarray*}
\int\phi^*_{2\nu}({\bf r})\phi_{2\nu}({\bf r})d{\bf r}=
{({\bf w}\tilde{\bf w})\over({\bf u}\tilde{\bf u})}.
\end{eqnarray*}

Now we multiply these orbitals by the spin-isospin functions and
construct the Slater determinant $\Psi$ of the nucleus $^{10}$Be.
A second Slater determinant $\Psi^*$ is constructed on the
conjugated functions. It can be considered as a result of rotation
of the coodinate frame transforming the vectors ${\bf u},{\bf w}$
into $\tilde{\bf u}, \tilde{\bf w}$. Therefore, the forthconing
procedure is nothing else but a Peierls--Yocozz method of the
angular momentum projection\cite{PY}. The overlap kernel (the
result of integration of the product of two Slater determinants
over all single-particle vectors ${\bf r}_i,i=1\dots10$) is
expanded over the Wigner's spherical functions. The kernels of
different operators are convenient to use instead of the wave
functions since the number of variables is reduced.

The overlap integral is actually very simple expression,
\begin{eqnarray}
\label{a4}
\int\Psi^*\Psi d{\bf r}_1d{\bf r}_2\cdot\cdot\cdot d{\bf r}_{10}=
I(^{10}\mbox{Be})=
({\bf u}\tilde{\bf u})^2({\bf w}\tilde{\bf w})^2.
\end{eqnarray}

The nucleus $^{10}$Be has two protons in its $p$-shell, and the
symmetry indices ($\lambda,\mu$) of the LR are $\lambda=2,\mu=2$.
It follows the relation (\ref{a4}), which can be rewritten in
another form due to the fact that the vectors involved are of unit
length,
\begin{eqnarray}
\label{a5}
I(^{10}\mbox{Be})=({\bf u}\tilde{\bf u})^2({\bf w}\tilde{\bf w})^2=
d_{11}^2d_{22}^2,
\end{eqnarray}
where $d_{11},~d_{22}$ are the elements of the rotation matrix.
These elements depend on the Euler angles only, so that $({\bf
u}\tilde{\bf u})=d_{11},~({\bf w}\tilde{\bf w})= d_{22}.$ The
Elliott's basis appears if the overlap (\ref{a5}) is expanded over
the Wigner's $D$-functions depending on the same Euler angles.
This expansion is a necessary element if the Peierls--Yocozz
method is implemented, because the weights of different angular
momentum states have to be found.

The overlap integral for a representation $(\lambda,\mu)$ is
\begin{eqnarray}
I(\lambda,\mu)=d_{11}^\lambda d_{22}^\mu.
\end{eqnarray}
Finding the coefficients $C^L_{K,\tilde{K}}$ of its expansion over
the $D$-functions complete the analysis of this expression. First
we write this expansion as follows,
\begin{eqnarray}
\label{z1} d_{11}^\lambda d_{22}^\mu=
\sum_L\sum_{K,\tilde{K}}C^L_{K,\tilde{K}}D^L_{K,\tilde{K}}.
\end{eqnarray}
Evidently, $C^L_{K,\tilde{K}}=C^L_{\tilde{K},K}$, so that the
expansion matrix is Hermitian.

One can see which $D$-functions enter the expansion (\ref{z1}) by
invoking the concept of the point group D$_2$ \cite{Dav}, elements
of which alternate the sign of one of the vectors ${\bf u,w}$
($\tilde{\bf u},\tilde{\bf w}$), or the vector orthogonal to both
of them. If the indices $\lambda$ and $\mu$ are even, the
expression $({\bf u}\tilde{\bf u})^\lambda({\bf w}\tilde{\bf
w})^\mu$ is invariant with respect to those transformations.
Therefore it may contain only those $D$-functions which belong to
the symmetric representation of D$_2$,
$$D^0_{0,0},~~D^2_{0,0},~~D^2_{2+,0},~~D^2_{0,2+},~~D^2_{2+,2+},~~D^3_{2-,2-}, ~~\mbox{etc.}$$
(ref. \cite{Dav}).

Yet not all of these functions may enter the expansion. Its actual
content depends on the choice of the coordinate frames used. The
number of basis states does not depend on this choice, but the
structure of the states does. We illustrate this on the example of
the overlap integral (\ref{a5}). Following Elliott, we direct the
$z$ ($\tilde{z}$) axis along the vector ${\bf u}$ ($\tilde{\bf
u}$), and $x$ ($\tilde{x}$) along ${\bf w}$ ($\tilde{\bf w}$).
Then the expansion takes the form
\begin{eqnarray*}
d_{11}^2d_{22}^2={2\over15}D^0_{0,0}+{5\over21}\left({1\over4}D^2_{0,0}+
{\sqrt{3}\over4}D^2_{0,2+}+{\sqrt{3}\over4}D^2_{2+,0}+
{3\over4}D^2_{2+,2+}\right)+
\end{eqnarray*}
\begin{eqnarray*}
+{1\over3}\left({3\over4}D^2_{0,0}-
{\sqrt{3}\over4}D^2_{0,2+}-{\sqrt{3}\over4}D^2_{2+,0}+
{1\over4}D^2_{2+,2+}\right)+{1\over6}D^3_{2-,2-}+
\end{eqnarray*}
\begin{eqnarray}
\label{a6}
+{9\over70}\left({4\over9}D^0_{0,0}-{2\sqrt{5}\over9}D^4_{0,2+}-
{2\sqrt{5}\over9}D^4_{2+,0}+{5\over9}D^4_{2+,2+}\right).
\end{eqnarray}
The five coefficients at the $D$-functions are the weight factors
of the basis states, and their sum equals 1 as it should be.
Moreover, the expansion (\ref{a6}) means that the $D$-functions
are maps of the basis states $\psi^\alpha_{Lm}$, where $m$ is the
projection of the orbital momentum on the external axis, $\alpha$
is an additional quantum number which may be necessary. Thus, for
the ground state of $^{10}$Be, $\psi_{0}=1$ with the weight
$2/15$, and for the state with $L=3$, $\psi_{3m}=D^3_{2-,m}$ with
the weight $1/6$. One of the states with $L=2$ has the weight
$5/21$ and the map
$$\psi^1_{2m}={1\over2}D^2_{0,m}+{\sqrt{3}\over2}D^2_{2+,m},$$
while the other, with the weight $1/3$,
$$\psi^2_{2m}={\sqrt{3}\over2}D^2_{0,m}-{1\over2}D^2_{2+,m}.$$
Finally, the state with $L=4$ has the weight $9/70$ and the map
$$\psi_{4m}={2\over3}D^0_{0,m}-{\sqrt{5}\over3}D^4_{2+,m}.$$
The wave functions obtained here are identical to those of the
non-axial rotator model\cite{DF} in the case when the non-axiality
parameter $\gamma=\pi/6.$ Thus we have established a link between
these two models which are based on essentially different
suggestions.

The overlap integral, both in the general (\ref{z1}) and in the
particular (\ref{a5}) cases, is also a density matrix calculated
by integrating of the product of two Slater determinants over all
single-particle variables. Unlike in the standard density matrix
\cite{Landau} where the integration is performed over some of the
single-particle variables while the remaining ones become
independent parameters, here the independent parameters are the
vectors ${\bf u},{\bf w}$ ($\tilde{\bf u},\tilde{\bf w}$). In
other words, the introduction of the density matrix is accompanied
by a transition from the space of single-particle variables to the
space of Euler angles. In this new space the basis states of SU(3)
representations are elegantly expressed as the Wigner's spherical
functions.

The density matrix is diagonal in the basis defined above. At the
same time, the basis functions differ in their weight factors.
That is why the density matrix describes a mixed state of the
nuclear system, because otherwise the weight factors would be
equal for all the functions.

Let $\lambda$ and $\mu$ be even. Then the following expansion
holds,
\begin{eqnarray}
\label{z2} I(\lambda,\mu)=\sum_L \sum_\alpha
w^\alpha_L(\lambda,\mu)\sum_m\psi^\alpha_{Lm}
\tilde{\psi}^\alpha_{Lm},
\end{eqnarray}
where $w^\alpha_L(\lambda,\mu)$ are the weight factors satisfying
the condition
\begin{eqnarray}
\sum_L \sum_\alpha w^\alpha_L(\lambda,\mu)=1.
\end{eqnarray}
If $\lambda\geq\mu$ and $L$ is even, then
\begin{eqnarray}
\label{y1} \psi^\alpha_{Lm}=\sum_{K=0}^\mu C^\alpha_{LK}
D^L_{K+,m}.
\end{eqnarray}
If $L$ is odd,
\begin{eqnarray}
\label{y2} \psi^\alpha_{Lm}=\sum_{K=0}^\mu C^\alpha_{LK}
D^L_{K-,m}.
\end{eqnarray}
In any case,
\begin{eqnarray}
\sum_{K=0}^\mu
C^\alpha_{LK}C^{\alpha'}_{LK}=\delta_{\alpha,\alpha'}.
\end{eqnarray}
In particular,
\begin{eqnarray}
w_0(\lambda,\mu)={(\lambda-1)!!(\mu-1)!!(\lambda+\mu)!!\over
\lambda!!\mu!!(\lambda+\mu+1)!!},
\end{eqnarray}
\begin{eqnarray}
w^1_2(\lambda,\mu)=
w_0(\lambda,\mu)~{5\over2}~{\lambda(\lambda+2)+\mu(\mu+2)
\over(\lambda+2)(\mu+2)(\lambda+\mu+3)}
%\times
%\end{eqnarray*}
%\begin{eqnarray}
%\times
\left\{{2\over\sqrt{3}}~{\lambda\mu+\lambda+\mu\over{\lambda-\mu}}y-
\sqrt{1+y^2}\right\},
\end{eqnarray}
\begin{eqnarray}
C^1_{20}=\sqrt{{\sqrt{1+y^2}+1\over2\sqrt{1+y^2}}},~~
C^1_{22}=\sqrt{{\sqrt{1+y^2}-1\over2\sqrt{1+y^2}}},
\end{eqnarray}
where
$$y=\sqrt3{{(\lambda+\mu+2)(\lambda-\mu)\over{\lambda(\lambda+2)+\mu(\mu+2)}}}.$$
\begin{eqnarray}
w^2_2(\lambda,\mu)=
w_0(\lambda,\mu)~{5\over2}~{\lambda(\lambda+2)+\mu(\mu+2)
\over(\lambda+2)(\mu+2)(\lambda+\mu+3)}
%\times
%\end{eqnarray*}
%\begin{eqnarray}
%\times
\left\{{2\over\sqrt{3}}~{\lambda\mu+\lambda+\mu\over{\lambda-\mu}}y+
\sqrt{1+y^2}\right\},
\end{eqnarray}
\begin{eqnarray}
C^2_{20}=\sqrt{{\sqrt{1+y^2}-1\over2\sqrt{1+y^2}}},~~
C^1_{22}=-\sqrt{{\sqrt{1+y^2}+1\over2\sqrt{1+y^2}}}.
\end{eqnarray}
The actual number of terms in (\ref{z2}) is defined by the
Elliott's rule for the basis states of an irreducible
representation of the SU(3) group.

Elliott's choice of the axes simplifies the classification of the
basis states. However, its drawback is a false impression that the
Elliott's scheme contradicts the experimental evidence related to
the structure of the wave functions. In the case of $^{10}$Be,
there are two D$_2$-symmetric states with $L=2$, and in neither of
them the projection of the angular momentum on an internal axis
$K$ is a constant of motion, although experiments show that the
rotational states are grouped in bands $K=0,~K=2$, etc.

In order to solve this contradiction, we redirect the axis $z$
($\tilde{z}$) along the vector orthogonal to ${\bf u}$ and ${\bf
w}$ ($\tilde{\bf u}$ and $\tilde{\bf w}$). This results in a
different structure of the expansion with the same weight factors,
\begin{eqnarray*}
d_{11}^2d_{22}^2={2\over15}D^0_{0,0}+{5\over21}D^2_{0,0}+
{1\over3}D^2_{2+,2+}+{1\over6}D^3_{2-,2-}+
\end{eqnarray*}
\begin{eqnarray}
\label{a7}
+{9\over70}\left({35\over36}D^4_{4+,4+}-{\sqrt{35}\over36}D^4_{4+,0}
-{\sqrt{35}\over36}D^4_{0,4+}+{1\over36}D^0_{0,0}\right).
\end{eqnarray}
One can see here that this choice of the rotation axis conserves
$K$ if $L=2$ which makes a better agreement between the Eliott's
scheme and the experimental data. As for the state with $L=4$, its
main component $K=4$ enters with the weight factor $35/36$, the
component $K=2$ is missing, and the component $K=2$ has a small
weight of $1/36$. Such a drastic difference in the structure of
states $L=2$ and $L=4$ may be related to an experimentally
observed phenomenon; in the first rotational band, the value of
$K$ undergoes a sudden change when $L$ reaches a critical value.

In general, the best choice of the rotation axis may be made
according to the following rules, depending on the SU(3) symmetry
indices ($\lambda,\mu$) of the LR.

\begin{enumerate}

\item If the indices of the LR are ($\lambda,0$), the overlap
integral is
$$({\bf u}\tilde{\bf u})^\lambda,$$
and the rotation axis should be directed along the vector ${\bf
u}$ ($\tilde{\bf u}$). Then, all allowed states have $K=0$. If
$\mu\ll\lambda$, components with $K\neq0$ appear, but their
amplitude is small in the states of the main rotational band, so
that they can be approximately treated as $K=0$ states.

\item Similarly, if the indices of the LR are ($0,\mu$), the overlap integral is
$$({\bf w}\tilde{\bf w})^\mu,$$
and the rotation axis should be directed along ${\bf w}$
($\tilde{\bf w}$). Again, all the states have $K=0$ only, and if
$\lambda\ll\mu$, $K$ may be considered approximately equal to
zero.

\item In the $\lambda=\mu$ case, the rotation axis should be
directed along $[{\bf u} \times {\bf w}]$ ($[\tilde{\bf u} \times
\tilde{\bf w}]$). If $\lambda$ and $\mu$ are slightly different, a
small admixture of $K=2$ states appear in the main rotational band
even at $L=2$, and $K=0$ admixtures are found in the $K=2$ band.

\end{enumerate}

\section{Overlap integral with the Hamiltonian}

The next stage in the realization of the Elliott's scheme for the
$^{10}$Be spectrum is the calculation of the overlap integral of
the generating Slater determinant with the Hamiltonian
$\hat{H}=\hat{T}+\hat{U}$, where $\hat{U}$ is a central exchange
potential having a Gaussian form,
\begin{eqnarray}
\label{a8}
\langle\Psi|\hat{U}|\Psi\rangle=A(I_1+I_2)+BI_3+(P-2A)I,
\end{eqnarray}
where
\begin{eqnarray*}
I_1=u^2u^{*2}({\bf ww}^*)^2,~~I_2=w^2w^{*2}({\bf uu}^*)^2,~~
I_3=([{\bf u}\times{\bf u}^*][{\bf w}\times{\bf w}^*])({\bf
uu}^*)({\bf ww}^*).
\end{eqnarray*}
Below we leave only those terms which split the spectrum and,
therefore, are of interest. The coefficients $A$, $B$ and $P$ are
expressed through amplitudes of the even components $V_{31}$ and
$V_{13}$ of the nucleon-nucleon potential, and
$$z^{-1}=1+{2r_0^2\over b_0^2},$$
where $r_0$ is the oscillator length, $b_0$ is the range of the
potential.
\begin{eqnarray}
A=z^{3/2}\left({1-z\over2}\right)^2~V_{13},~~
B=z^{3/2}\left({1-z\over2}\right)^2~(V_{13}+3V_{31}).
\end{eqnarray}
If we direct the axis $z$ ($\tilde{z}$) along $[{\bf u} \times
{\bf w}]$, ($[\tilde{\bf u}\times \tilde{\bf w}]$),  we obtain
\begin{eqnarray*}
I_1+I_2={2\over3}D^0_{0,0}+{1\over3}D^2_{0,0}+ D^2_{2+,2+},
\end{eqnarray*}
\begin{eqnarray*}
I_3={1\over30}D^0_{0,0}+{2\over21}D^2_{0,0}-
%\end{eqnarray*}
%\begin{eqnarray*}
{9\over70}\left({35\over36}D^4_{4+,4+}-{\sqrt{35}\over36}D^4_{4+,0}
-{\sqrt{35}\over36}D^4_{0,4+}+{1\over36}D^0_{0,0}\right).
\end{eqnarray*}

It is easy now to find the energies of the five states of
$^{10}$Be. The ground $0^+$ state is located at
\begin{eqnarray}
E_0=5A+{1\over4}B.
\end{eqnarray}
$E_0$ will be used as a reference point for the excited states,
\begin{eqnarray}
\label{v1}
E^1_2-E_0=-{18\over5}\left(A-{B\over4}\right)-{3\over4}B,
\end{eqnarray}
\begin{eqnarray}
\label{v2} E^2_2-E_0=-2\left(A-{B\over4}\right)-{3\over4}B,
\end{eqnarray}
\begin{eqnarray}
E_3-E_0=-5\left(A-{B\over4}\right)-{3\over2}B,
\end{eqnarray}
\begin{eqnarray}
E_4-E_0=-5\left(A-{B\over4}\right)-{5\over2}B.
\end{eqnarray}

Note that if the Volkov potential is used, the states $L=2,K=0$
and $L=2,K=2$ are degenerate, simply because this potential has
$V_{31}=V_{13}$. Fig.~1 shows the spectrum of $^{10}$Be for the
Volkov potential with $m=0.6$, if $r_0=1.64$ fm. This value of
$r_0$ is optimized by variation; it minimizes the energy of the
ground state and correspond to the r.m.s. radius 2.29 fm (the
experimental value is $2.3\pm0.2$ fm\cite{Tan}).

\begin{figure}
\begin{center}
\includegraphics[width=8cm]{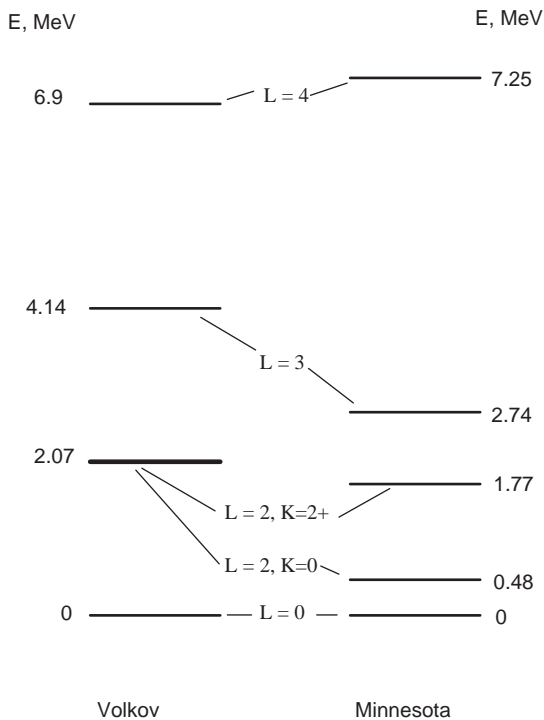}
\end{center}
\caption{Spectrum of $^{10}$Be for the Hamiltonian (\ref{a8}) with
Volkov and Minnesota interactions.}
\end{figure}

While the Volkov potential is consistent with the observed value
for the radius of $^{10}$Be, the Minnesota potential with $u=0.98$
(this value of the exchange mixture parameter seems to be the best
choice for the nuclei in the first half of the $p$-shell) leads to
$r_0$=1.43 fm, and besides, yields a wrong sequence of excited
states. The saturation is not reached for both potentials but,
again, this is more evident in the Minnesota case (light nuclei
being considered). Nevertheless, if $r_0$ is set to $1.64$ fm in
the Minnesota case, one obtains the spectrum shown in Fig.~1 on
the right. Now the $2^+$ states split as $V_{31}\neq V_{13}$.
However, the splitting is just 1.29 MeV, which is twice smaller
than observed (2.59 MeV). The splitting $E_2^2-E_2^1$ is
proportional to $V_{31}-V_{13}$. Therefore, this value has to be
increased in order to achieve a good description of the
experimental results within the model. This suggestion is also
supported by the fact that the value
$${1\over2}(E_2^2+E_2^1)-E_0=1.125 \, \mbox{MeV}$$
appears to be four times less than the experimental one, and
almost twice less than in the Volkov potential case ($2.07$ MeV).

Historically, both potentials were designed to describe the
experimental data for the nuclei in the beginning of the
$p$-shell. It is therefore not surprising that the their
application to heavier nuclei may be accompanied by some
corrections in the potentials themselves.

The Elliott's scheme predicts the existence of the levels $3^+$
and $4^+$. Energy of the former is close to the sum of the
energies of the $2^+$ states. The wave function of the latter has
predominantly $K=4$ which affects the probabilities of E2
transitions. The isoscalar electric transition from the $4^+$
state to the $L=2,~K=0$ state is forbidden, leaving allowed only
the transition to the $L=2,~K=2$ state.

It is important that a right choice of the central nucleon-nucleon
interaction makes the states $L=2,K=0$ and $L=2,K=2$ uncoupled.

\section{Effective Hamiltonian of the SU(3) model}

When the overlap integral (\ref{a8}) was calculated, the operator
$\hat{H}$ was acting on the generating Slater determinant $\Psi$.
Its SU(3) symmetry was (2,2), and the result was projected to the
states with the same symmetry. The projection was performed by an
integration. Therefore, the overlap integral (\ref{a8}) keeps us
within the basis of the representation $(2,2)$. In order to
understand how the Hamiltonian $\hat{H}$ acts in this
representation, we define an operator $\hat{H}({\bf u,w})$ as
follows,
\begin{eqnarray}
\langle\Psi|\hat{H}|\Psi\rangle=\hat{H}({\bf
u,w})\langle\Psi|\Psi\rangle.
\end{eqnarray}
All operators which acting onto the basis states of an irreducible
representations produce only those states which belong to the same
representation are known: they are the group generators. We are
interested here only in those expressions constructed from the
SU(3) generators which are scalars with respect to the O(3) group.
They are known, too \cite{DW1}, \cite{DW2}. They are the operator
${\bf QQ}$ introduced by Elliott, and the operators
$\hat{\Omega}={\bf LQL}$ and $\hat{\Omega}_1={\bf LQQL}$
\cite{BM}. All other operators may be written as polynomials of
these three.

In our model the wave functions are represented by superpositions
of the $D$-functions. Therefore, as in the case of the Hamiltonian
of the triaxial rotator, all operators may be expressed via the
SU(3) symmetry indices $\lambda,\mu$ and the three projections of
the orbital momentum $L$ to the internal axes:
$\hat{L}_1,\hat{L}_2,\hat{L}_3$. It has been shown\cite{FD} that
\begin{eqnarray}
\label{z3}
-\hat{\Omega}=a\hat{L}^2_1+b\hat{L}^2_2+c\hat{L}^2_3+\hat{\Omega}_2,
\end{eqnarray}
\begin{eqnarray}
-\hat{\Omega}_1=a^2\hat{L}^2_1+b^2\hat{L}^2_2+
c^2\hat{L}^2_3-{2\over3}\hat{\bf
L}^2-\hat{L}^2_2+{1\over2}(\hat{L}^2_1\hat{L}^2_3+
\hat{L}^2_3\hat{L}^2_1)+2b\hat{\Omega}_2,
\end{eqnarray}
where
\begin{eqnarray}
a={2\lambda+\mu+3\over3},~~b={-\lambda+\mu\over3},~~c={-\lambda-2\mu-3\over3}.
\end{eqnarray}
\begin{eqnarray}
\hat{\Omega}_2={i\over6}\left(\hat{L}_1\hat{L}_2\hat{L}_3+
\hat{L}_2\hat{L}_3\hat{L}_1+\hat{L}_3\hat{L}_1\hat{L}_2+
%\end{eqnarray*}
%\begin{eqnarray}
%\left.+
\hat{L}_2\hat{L}_1\hat{L}_3+\hat{L}_3\hat{L}_2\hat{L}_1+
\hat{L}_1\hat{L}_3\hat{L}_2\right).
\end{eqnarray}

The structure of the operators $\hat{\Omega}$ and $\hat{\Omega}_1$
is similar to that of the Hamiltonian of the triaxial rotator, but
the inertial parameters (the coefficients of
$\hat{L}^2_1,~\hat{L}^2_2,~ \hat{L}^2_3$) are not the principle
values of the tensor of inertia, but rather the principle values
of the tensor of the intrinsic quadrupole momentum (in the case of
$\hat{\Omega}$) or their squared values (in the case of
$\hat{\Omega}_1$). Additional terms
$$\hat{\Omega}_2~~\mbox{and}~~{1\over2}(\hat{L}^2_1\hat{L}^2_3+
\hat{L}^2_3\hat{L}^2_1)$$ only lead to the finite number of basis
functions of an SU(3) representation whereas in the triaxial
rotator case the basis is infinite.

The operators $\hat{\Omega}$ and $\hat{\Omega}_1$ commute with
$\hat{\bf L}^2$, therefore the orbital momentum $L$, as expected,
is a constant of motion. Besides, these operators are
D$_2$-invariant, so that their eigenfunctions have a definite
D$_2$ symmetry. However, among all eigenfunctions of
$\hat{\Omega}$ and $\hat{\Omega}_1$, only those are of physical
interest which have a D$_2$ symmetry identical to that of the
basis states produced by the overlap integral (\ref{z1}).

It is natural to seek the eigenfunctions $\phi_{L,m}$ of
$\hat{\Omega}$ (or $\hat{\Omega}_1$, or a linear combination of
$\hat{\Omega}$ and $\hat{\Omega}_1$) in the form of a
superposition of the functions
$$\sqrt{w_L^\alpha}\psi^{\alpha}_{L,m},$$
taking into account their normalization, that is
\begin{eqnarray}
\phi_{L,m}=\sum_{\alpha}B_{L,\alpha}\sqrt{w_L^\alpha}\psi^{\alpha}_{L,m}.
\end{eqnarray}
Then the coefficients $B_{L,\alpha}$ must satisfy the following
set of equations,
\begin{eqnarray}
\sum_{\alpha'}(\langle L,\alpha|\hat{\Omega}|L,\alpha'\rangle-
\delta_{\alpha,\alpha'})B_{L,\alpha'}=0.
\end{eqnarray}
One may calculate the matrix elements $\langle
L,\alpha|\hat{\Omega}|L,\alpha'\rangle$, by calculating first the
matrix element of $\hat{\Omega}$ between the generating functions
and then expanding the result,
\begin{eqnarray}
\langle\Psi|\hat{\Omega}|\Psi\rangle=\sum_L \sum_{\alpha,\alpha'}
\sqrt{w^\alpha_L(\lambda,\mu)w^{\alpha'}_L(\lambda,\mu)} \langle
L,\alpha|\hat{\Omega}|L,\alpha'\rangle
\sum_m\psi^\alpha_{Lm}\tilde{\psi}^{\alpha'}_{Lm}.
\end{eqnarray}
Alternatively, a simple formula (\ref{z3}) for $\hat{\Omega}$
provides for another way,
\begin{eqnarray}
\hat{\Omega}\sqrt{w_L^\alpha}\psi^{\alpha}_{L,m}=
\sum_{\alpha'}\langle
L,\alpha|\hat{\Omega}|L,\alpha'\rangle\sqrt{w_L^{\alpha'}}
\psi^{\alpha}_{L,m}.
\end{eqnarray}
The action of the operators $\hat{L}_1,\hat{L}_2,\hat{L}_3$ on the
$D$-functions is known (cf. \cite{Landau}). Both approaches lead
to the same Hermitian matrix,
$$||\langle L,\alpha|\hat{\Omega}|L,\alpha'\rangle||.$$
Diagonalization of this matrix yields the eigenvalues of $\Omega$.
At a given $L$, the size of the matrix depends on the number of
possible values of $\alpha$.

Finally, one more possibility is to expand $\phi_{L,m}$ over the
$D$-functions straightforwardly,
\begin{eqnarray}
\label{z4}
\phi_{L,m}=N_L\sum_K A_{L,K}D^L_{K\pm,m}.
\end{eqnarray}
This way is the simplest, it results in a non-Hermitian matrix
\begin{eqnarray}
\label{z5}
||\langle L,K\pm|\hat{\Omega}|L,K'\pm\rangle||,
\end{eqnarray}
because $\hat{\Omega}_2$ (a term in both $\hat{\Omega}$ and
$\hat{\Omega}_1$) contains an imaginary unit $i$ as a factor, and
that makes it a non-self-conjugate operator. This does not mean,
however, that the matrix will necessarily have complex eigenvalues
\footnote{A Hermitian matrix has real eigenvalues only, but a
reverse statement is not right, in general.}. The coefficients
$A_{L,K}$ are found from the set
\begin{eqnarray}
\sum_{K'}(\langle
L,K\pm|\hat{\Omega}|L,K'\pm\rangle-\delta_{K,K'})A_{L,K}=0,
\end{eqnarray}
where the matrix elements $\langle
L,K\pm|\hat{\Omega}|L,K'\pm\rangle$ are defined by
\begin{eqnarray}
\hat{\Omega}D^L_{K\pm,m}= \sum_{K'}\langle
L,K\pm|\hat{\Omega}|L,K'\pm\rangle D^L_{K'\pm,m}.
\end{eqnarray}
The normalization factor $N_{L}$ of the eigenvalues (\ref{z4}) is
found from the condition
\begin{eqnarray}
\label{z6} N_L\sum_\alpha{1\over\sqrt{w_L^\alpha}}\sum_K
A_{L,K}C^\alpha_{L,K}=1.
\end{eqnarray}
The coefficients $C^\alpha_{L,K}$ were defined above ((\ref{y1})
and (\ref{y2})). They can be found by expanding the overlap
integral (\ref{z1}) over the basis states of an irreducible
representation of the SU(3) group. The summation over $K$ in
(\ref{z6}) signifies the projection of an eigenfunction of
$\hat{\Omega}$ to the allowed states. Hence those eigenfunctions
of $\hat{\Omega}$ for which each of the sums
$$\sum_K A_{L,K}C^\alpha_{L,K}$$
vanishes are forbidden. They do not satisfy the Pauli exclusion
principle and must be disregarded. With $L$ increasing, such
states appear, sooner or later, and this must be taken into
account in the calculations of spectra.

\section{General remarks on the effective Hamiltonian}

At least at the first stage, the effective Hamiltonian is better
to be constructed as a linear combination of the operators ${\bf
QQ}$, $\hat{\Omega}$ and $\hat{\Omega}_1$. It is now clear that
the result will be, in fact, the Hamiltonian of the triaxial
rotator model with some additional terms which may be important in
the actual calculations of the LR spectra. The coefficients of the
three operators may then serve as phenomenological parameters.

The question is, can such a Hamiltonian describe the observed
low-energy spectra of even-even nuclei? One may believe that the
phenomenological parameters would be the same for a whole range of
nuclei while the SU(3) symmetry indices would change from nucleus
to nucleus according to known general rules.

Such an approach has been used in \cite{DW1} and \cite{DW2}, and
there an attempt was made to extend the model to excited states
with higher $L$. Then, one may question, whether it suffices to
have the three operators entering the Hamiltonian in their first
order, or a higher-order terms are necessary?

In Ref. \cite{OSt} the authors studied the $^{20}$Ne and $^{44}$Ti
nuclei and concluded, that the effective Hamiltonian of their LR
$(\lambda,0)$ ((8,0) for $^{20}$Ne and (12,0) for $^{44}$Ti),
derived from the overlap integrals of the generating Slater
determinants with the Gaussian interaction, contains first two
orders of ${\bf QQ}$ for $^{20}$Ne, and first three orders for
$^{44}$Ti. It became clear that higher orders are necessary for
higher oscillator shells. The influence of higher-order terms is
increasing with $L$. For representations $(\lambda,0)$, the
operators $\hat{\Omega}$ and $\hat{\Omega}_1$ degenerate and
reduce to the operator $\hat{\bf L}^2$. Hence to derive the terms
containing high orders of $\hat{\Omega}$ and $\hat{\Omega}_1$, one
has to use the LRs with $\lambda\ne0, \mu\ne0$ and go beyond the
$p$-shell. Alternatively, the higher-order terms may be included
in the Hamiltonian phenomenologically.

\section{The mass quadrupole operator}

We now show that $a,~b,~c$ are the principle values of the
quadrupole momentum. To start with, the explicit form of the
operator of the mass quadrupole momentum $\hat{Q}_{\alpha\beta}$
in the intrinsic coordinate frame is shown in \cite{FD}. The same
operator in the laboratory frame (with our choice of axes) is
\begin{eqnarray}
\label{z7} \hat{Q}_{2m}={1\over3}(-\lambda+\mu)D^2_{m,0}+
{\sqrt{3}\over3}(\lambda+\mu)D^2_{m,2+}
%\end{eqnarray*}
%\begin{eqnarray}
-{\sqrt{3}\over3}\left(D^2_{m,1+}\hat{L}_1+iD^2_{m,1-}\hat{L}_3+
D^2_{m,2-}\hat{L}_2\right).
\end{eqnarray}
Here
$${1\over3}(-\lambda+\mu)~~\mbox{and}~~{\sqrt{3}\over3}(\lambda+\mu)$$
are the principle values of the traceless tensor of the quadrupole
momentum.

It follows from (\ref{z7}) that
\begin{eqnarray}
{\bf QQ}={2\over3}(\lambda^2+\lambda\mu+\mu^2+3\lambda+3\mu)-
{1\over2}\hat{\bf L}^2.
\end{eqnarray}

Also note that
\begin{eqnarray*}
-{\sqrt{3}\over3}\left(D^2_{m,1+}\hat{L}_1+iD^2_{m,1-}\hat{L}_3+
D^2_{m,2-}\hat{L}_2\right)D^2_{m',2+}=
\end{eqnarray*}
\begin{eqnarray}
=-{\sqrt{3}\over3}\left(D^2_{m,1+}D^2_{m',1+}-D^2_{m,1-}D^2_{m',1-}
+2D^2_{m,2-}D^2_{m',2-}\right).
\end{eqnarray}
We have arrived to an expected, D$_2$-invariant expression.
Indeed, acting on such an expression, the operator of the
quadrupole momentum must yield another D$_2$-invariant expression.

Consider now the Eliiott's choice of axes.
\begin{eqnarray*}
\hat{Q}_{2m}={2\over3}\left\{(\bar{a}-{\bar{b}+\bar{c}\over2})D^2_{m,0}+
{\sqrt{3}\over2}(\bar{b}-\bar{c})D^2_{m,2+}\right\}+...
%\end{eqnarray*}
%\begin{eqnarray}
={1\over3}(2\lambda+\mu)D^2_{m,0}+ {\sqrt{3}\over3}\mu
D^2_{m,2+}+...
\end{eqnarray*}
By definition,
\begin{eqnarray*}
\bar{a}={1\over3}(2\lambda+\mu)=a-1,~~\bar{b}={1\over3}(-\lambda+\mu)=b,~~
\bar{c}={1\over3}(-\lambda-2\mu)=c+1.
\end{eqnarray*}
Then, for the representation $(\lambda,0)$ one obtains
\begin{eqnarray}
\hat{Q}_{2m}={2\lambda\over3}D^2_{m,0}+...
\end{eqnarray}
In this case, the axially symmetric nucleus has a prolate shape
and therefore its intrinsic quadrupole momentum\footnote{The
quadrupole momentum is a traceless tensor with two main
components, $Q_{20}$ and $Q_{22+}$. In the axially-symmetric case
the second component vanishes, and the quadrupole momentum is said
to be oriented along one of its principal axes.} is directed along
the rotation axis and is positive.

Yet another choice,
\begin{eqnarray*}
\hat{Q}_{2m}={2\over3}\left\{(\bar{c}-{\bar{a}+\bar{b}\over2})D^2_{m,0}+
{\sqrt{3}\over2}(\bar{a}-\bar{b})D^2_{m,2+}\right\}+... =
\end{eqnarray*}
\begin{eqnarray*}
=-{1\over3}(2\mu+\lambda)D^2_{m,0}+{\sqrt{3}\over3}\lambda
D^2_{m,2+}+...
\end{eqnarray*}
is convenient for the limiting case of $(0,\mu)$, when
\begin{eqnarray}
\hat{Q}_{2m}=-{2\mu\over3}D^2_{m,0}+...
\end{eqnarray}
Now the nucleus has an oblate shape, the quadrupole momentum is
again oriented along the rotation axis but this time, negative.

Finally, we show the matrix elements of the isoscalar E1
transition from the ground state to the states $\psi^\alpha_{2,m}$
of the Elliott's basis when the indices $\lambda,\mu$ are both
even.
\begin{eqnarray}
\langle L=2,\alpha=1,m|\hat{Q}_{2m}|L=0\rangle
%\end{eqnarray*}
%\begin{eqnarray}
=\sqrt{{w_0\over w^1_2}}\left\{{1\over3}(-\lambda+\mu)C^1_{20}+
{\sqrt{3}\over3}(\lambda+\mu)C^1_{22}\right\},
\end{eqnarray}
\begin{eqnarray}
\langle L=2,\alpha=2,m|\hat{Q}_{2m}|L=0\rangle=
%\end{eqnarray*}
%\begin{eqnarray}
\sqrt{{w_0\over w^2_2}}\left\{{1\over3}(-\lambda+\mu)C^2_{20}+
{\sqrt{3}\over3}(\lambda+\mu)C^2_{22}\right\}.
\end{eqnarray}

\section{Effective Hamiltonian for $^{10}$Be}

We shall now try to built an effective Hamiltonian for $^{10}$Be
from ${\bf L}^2,~\hat{\Omega},~\hat{\Omega}_1$. The only condition
we impose on the Hamiltonian is the equivalence of its spectra
(eigenvalues and eigenvectors) to that found above in the LR
approximation.

\begin{eqnarray}
\hat{H}({\bf u,w})=B\left({3\over2}-{1\over8}~{\bf L}^2\right)+
\left(A-{B\over4}\right)\left(p+q{\bf
L}^2+r\hat{\Omega}_1+s\hat{\Omega}\right),
\end{eqnarray}
where $p,q,r,s$ are the coefficients to be found.

We have noted earlier that the matrix elements of the norm and
Hamiltonian operators between the basis functions $\psi_{2m}^1$
and $\psi_{2m}^2$ with $L=2$ are diagonal only. Meanwhile, the
eigenfunctions of $\hat{\Omega}$ are linear combinations of these
states. Therefore, $s=0$, and it is the operator $\hat{\Omega}_1$
that splits the two levels with $L=2$.  It follows (\ref{v1}) and
(\ref{v2}), that
\begin{eqnarray}
E^2_2-E_2^1={8\over5}~\left(A-{B\over4}\right).
\end{eqnarray}
The eigenvalues of $\hat{\Omega}_1$ in these states are
\begin{eqnarray}
\hat{\Omega}_1\psi_{2m}^1=-44~\psi_{2m}^1,~~
\hat{\Omega}_1\psi_{2m}^2=-12~\psi_{2m}^2.
\end{eqnarray}
Using
\begin{eqnarray}
E_0=5A+{1\over4}B,~\mbox{and}~
E^1_2-E_0=-{18\over5}\left(A-{B\over4}\right)-{3\over4}B,
\end{eqnarray}
we define $p,q,r$ and obtain the effective Hamiltonian in its
final form,
\begin{eqnarray}
\hat{H}({\bf u,w})=B\left({3\over2}-{1\over8}~{\bf L}^2\right)+
\left(A-{B\over4}\right)\left(5-{7\over30}~{\bf
L}^2+{1\over20}~\hat{\Omega}_1 \right).
\end{eqnarray}
One may check straightforwardly that this Hamiltonian yields the
same values of $E_3$ and $E_4$, too.

In the phenomenological approach, the coefficients $A$ and $B$
should be chosen so that to reproduce the experimental spectrum of
$^{10}$Be. It would suffice then to use the energy of the first
three states, $E_0,E_2^1$ and $E_2^2$. The energies $E_3$ and
$E_4$ could serve as a test. But their experimental values are not
known at present.

\section{Summary}

We have shown that the realization of the SU(3) LR approximation
with the microscopic Hamiltonian and a nucleon-nucleon potential
(such as the Volkov and Minnesota potentials) may be reduced to
the calculations with an effective potential. The latter is a
linear combination of the operators $\hat{\bf L}^2$,
$\hat{\Omega}$ and $\hat{\Omega}_1$, as well as their higher
orders. In many ways this Hamiltonian is similar to that of the
triaxial rotator model. However, the inertial parameters of such a
rotator are not inversely proportional to the principal values of
the tensor of inertia, but proportional to the principal values of
the tensor of the quadrupole momentum. Both the basis functions of
an irreducible representation of the SU(3) group and the basis
functions of the rotator model are superpositions of Wigner's
$D$-functions, but their normalization should be calculated
separately.

Some problems appearing in the calculations of the spectra of
nuclei in the second half of the $p$-shell with the Volkov and
Minnesota potentials are discussed.

Some remarks on the structure of effective Hamiltonians of medium
and heavy nuclei are also given.

\end{document}